\documentclass[12pt]{llncs}
\usepackage{amsmath,amssymb,fullpage,graphicx,mathptmx}

\begin{document}
\title{Fixed point theorem and aperiodic tilings\thanks{Partially
       supported by ANR (Sycomore and Nafit grants) and RFBR (05-01-02803,
       06-01-00122a).}}
\author{Bruno Durand\inst{1},
Andrei Romashchenko\inst{1,2}, Alexander Shen\inst{1,2}}
\institute{LIF, CNRS \& Univ. de Provence, Marseille
\and Institute for Information Transmission Problems, Moscow}
\date{}
\maketitle

\pagestyle{plain}

\begin{abstract}

We propose a new simple construction of an aperiodic tile set based on
self-referential (fixed point) argument.

\medskip

People often say about some discovery that it appeared ``ahead
of time'', meaning that it could be fully understood only in the
context of ideas developed later. For the topic of this note,
the construction of an aperiodic tile set based on the
fixed-point (self-referential) approach, the situation is
exactly the opposite. It should have been found in 1960s when
the question about aperiodic tile sets was first asked: all the
tools were quite standard and widely used at that time. However,
the history had chosen a different path and many nice geometric
\emph{ad hoc} constructions were developed instead (by Berger,
Robinson, Penrose, Ammann and many others, see~\cite{grunbaum};
a popular exposition of Robinson-style construction is given
in~\cite{intelligencer}). In this note we try to correct this
error and present a construction that should have been discovered first
but seemed to be unnoticed for more that forty years.

\end{abstract}

\section{The statement: aperiodic tile sets}

A tile is a square with colored sides. Given a set of tiles, we
want to find a tiling, i.e., to cover the plane by (translated
copies of) these tiles in such a way that colors match (a common
side of two neighbor tiles has the same color in both).\footnote{%
 Tiles appeared first in the context of \emph{domino problem}
 posed by Hao Wang.
 Here is the original formulation from~\cite{wang}:
 ``Assume we are given a finite set of square plates of the same size
  with edges colored, each in a different manner. Suppose further there are infinitely many
 copies of each plate (plate type). We are not permitted to rotate or reflect a plate. The
 question is to find an effective procedure by which we can decide, for each given finite set
 of plates, whether we can cover up the whole plane (or, equivalently, an infinite quadrant
 thereof) with copies of the plates subject to the restriction that adjoining edges must have
 the same color.''  This question (domino problem)
 is closely related to the existence of aperiodic tile sets:
 (1)~if they did not exist, domino problem would be decidable for some simple reasons (one
 may look in parallel for a periodic tiling or a finite region that cannot be tiled) and (2)~the
 aperiodic tile sets are used in the proof of the undecidability of domino problem. However,
 in this note we concentrate on aperiodic tile sets only.}

For example, if tile set consists of two tiles
\begin{figure}[h]
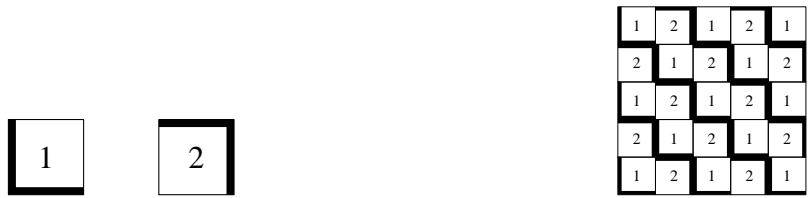

\begin{center}
\includegraphics[scale=1]{gurevich-4.mps}\hspace*{0.3\hsize}
\includegraphics[scale=1]{gurevich-5.mps}
\end{center}
\caption{Tile set that has only periodic tilings}
\end{figure}
(one has black lower and left side and white right and top sides, the other has the
opposite colors), it is easy to see that only periodic (checkerboard) tiling is possible.
However, if we add some other tiles the resulting tile set may admit also non-periodic
tilings (e.g., if we add all $16$ possible tiles, any combination of edge colors becomes
possible). It turns out that there are other tile set that have \emph{only} aperiodic
tilings.

Formally: let $C$ be a finite set of \emph{colors} and let $\tau\subset
C^4$ be a set of \emph{tiles}; the components of the quadruple
are interpreted as upper/right/lower/left colors of a tile. Our example tile set
with two tiles
is represented then as
     $$
\{\langle \text{white}, \text{white}, \text{black}, \text{black}\rangle,
\langle \text{black}, \text{black}, \text{white}, \text{white}\rangle\}.
     $$
A \emph{$\tau$-tiling} is a mapping $\mathbb{Z}^2\to \tau$ that
satisfies matching conditions. Tiling $U$ is called
\emph{periodic} if it has a \emph{period}, i.e., if there exists
a non-zero vector $T\in\mathbb{Z}^2$ such that $U(x+T)=U(x)$ for
all $x$.

Now we can formulate the result (first proven by
Berger~\cite{berger}):

\textbf{Proposition}. \emph{There exists a finite tile set
$\tau$ such that $\tau$-tilings exist but all of them are
aperiodic}.

There is a useful reformulation of this result. Instead of
tilings we can consider two-dimensional infinite words in some
finite alphabet $A$ (i.e., mappings of type $\mathbb{Z}^2\to A$)
and put some local constraints on them. This means that we
choose some positive integer $N$ and look at the word through a
window of size $N\times N$. Local constraint then says which
patterns of size $N\times N$ are allowed to appear in a window.
Now we can reformulate our Proposition as follows: \emph{there
exists a local constraint that is consistent \textup(some
infinite words satisfy it\textup) but implies aperiodicity
\textup(all satisfying words are aperiodic\textup)}.

It is easy to see that these two formulations are equivalent.
Indeed, the color matching condition is $2\times 2$ checkable.
On the other hand, any local constraint can be expressed in
terms of tiles and colors if we use $N\times N$-patterns as
tiles and $(N-1)\times N$-patterns as colors; e.g., the right
color of $(N\times N)$-tile is the tile except for its left
column; if it matches the left color of the right neighbor,
these two tiles overlap correctly.

\section{Why theory of computation?}

At first glance this proposition has nothing to do with theory
of computation. However, the question appeared in the context of
the undecidability of some logical decision problems, and, as we
shall see, can be solved using theory of computations. (A rare
chance to convince ``normal'' mathematicians that theory of
computations is useful!)

The reason why theory of computation comes into play is that
rules that determine the behavior of a computation device~---
say, a Turing machine with one-dimensional tape~--- can be
transformed into local constraints for the space-time diagram
that represents computation process. So we can try to prove the
proposition as follows: consider a Turing machine with a very
complicated (and therefore aperiodic) behavior and translate its
rules into local constraints; then any tiling represents a
time-space diagram of a computation and therefore is aperiodic.

However, this na\"\i ve approach does not work since local
constraints are satisfied also at the places where no computation happens (in the
regions that do not contain the head of a Turing machine) and
therefore allow periodic configurations. So a more sophisticated
approach is needed.

\section{Self-similarity}

The main idea of this more sophisticated approach is to
construct a ``self-similar'' set of tiles. Informally speaking,
this means that any tiling can be uniquely split by vertical and
horizontal lines into $M\times M$ blocks that behave exactly
like the individual tiles. Then, if we see a tiling and zoom out
with scale $1:M$, we get a tiling with the same tile set.

Let us give a formal definition. Assume that a non-empty set of tiles
$\tau$ and positive integer $M>1$ are fixed. A \emph{macro-tile}
is a square of size $M\times M$ filled with matching tiles from
$\tau$. Let $\rho$ be a non-empty set of macro-tiles.

\textbf{Definition}.
We say that $\tau$
\emph{implements} $\rho$ if any $\tau$-tiling can be uniquely
split by horizontal and vertical lines into macro-tiles from
$\rho$.

\medskip
Now we give two examples that illustrate this definition: one
negative and one positive.

\medskip
\textbf{Negative example}: Consider a set $\tau$ that consists
of one tile with all white sides. Then there is only one
macro-tile (of given size $M\times M$). Let $\rho$ be a
one-element set that consists of this macro-tile. Any
$\tau$-tiling (i.e., the only possible $\tau$-tiling) can be
split into $\rho$-macro-tiles. However, the splitting lines are
not unique, so $\tau$ does \emph{not} implements $\rho$.

\begin{figure}[h]
\begin{center}
\includegraphics[scale=1]{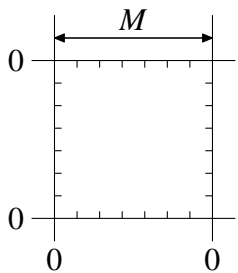}
\end{center}
\caption{The only element of $\rho$: border colors are pairs
         that contain $0$}
\label{2.mps}
\end{figure}

\medskip
\textbf{Positive example}: Let $\tau$ is a set of $M^2$ tiles
that are indexed by pairs of integers modulo $M$:
\begin{figure}[h] 	
\begin{center}
\includegraphics[scale=1]{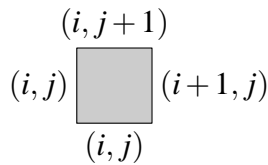} 	
\end{center}
\caption{Elements of $\tau$ (here $i,j$ are integers modulo
$M$)}
\label{1.mps}
\end{figure}
The colors are pairs of
integers modulo $M$ arranged as shown (Fig.~\ref{1.mps}).	
Then there exists only
one $\tau$-tiling (up to translations), and this tiling can be
uniquely split into $M\times M$ squares whose borders have
colors $(0,j)$ and $(i,0)$. Therefore, $\tau$ implements a set
$\rho$ that consists of one macro-tile (Fig.~\ref{2.mps}).

\medskip
\textbf{Definition}.
A set of tiles $\tau$ is \emph{self-similar} if it implements
some set of macro-tiles $\rho$ that is isomorphic to $\tau$.

This means that there exist a 1-1-correspondence between $\tau$
and $\rho$ such that matching pairs of $\tau$-tiles correspond
exactly to matching pairs of $\rho$-macro-tiles.

\medskip

The following statement follows directly from the definition:

\medskip

\textbf{Proposition}. \emph{A self-similar tile set $\tau$ has
only aperiodic tilings}.

\textbf{Proof}. Let $T$ be a period of some $\tau$-tiling $U$. By
definition $U$ can be uniquely split into $\rho$-macro-tiles.
Shift by $T$ should respect this splitting (otherwise we get a
different splitting), so $T$ is a multiple of $M$. Zooming the
tiling and replacing each $\rho$-macro-tile by a corresponding
$\tau$-tile, we get a $T/M$-shift of a $\tau$-tiling. For the
same reason $T/M$ should be a multiple of $M$, then we zoom out
again etc. We conclude therefore that $T$ is a multiple of $M^k$ for
any $k$, i.e., $T$ is a zero vector.\raisebox{-.6ex}{\hbox{\scriptsize $\square$}}

Note also that any self-similar set $\tau$ has at least one
tiling. Indeed, by definition we can tile a $M\times M$ square
(since macro-tiles exist). Replacing each $\tau$-tile by a
corresponding macro-tile, we get a $\tau$-tiling of $M^2\times
M^2$ square, etc. In this way we can tile an arbitrarily large
finite region, and then standard compactness argument (K\"onig's
lemma) shows that we can tile the entire plane.

So it remains to construct a self-similar set of tiles (a set of
tiles that implements itself, up to an isomorphism).

\section {Fixed points and self-referential constructions}
\label{fixed}

The construction of a self-similar tile set is done in two steps.
First (in Section~\ref{implementing}) we explain how to
construct (for a given tile set $\sigma$) another tile set $\tau$
that implements $\sigma$ (i.e., implements a set of macro-tiles
isomorphic to $\sigma$). In this construction the tile set $\sigma$
is given as a program $p_\sigma$ that checks whether four bit strings (representing
four side colors) appear in one $\sigma$-tile. The tile set $\tau$ then
guarantees that each macro-tile encodes a computation where
$p_\sigma$ is applied to these four strings (``macro-colors'')
and accepts them.

This gives us a mapping: for every $\sigma$ we have $\tau=\tau(\sigma)$
that implements $\sigma$ and depends on $\sigma$. Now we need a
fixed point of this mapping where $\tau(\sigma)$ is isomorphic to $\sigma$.
It is done (Section~\ref{itself}) by a classical self-referential trick that appeared as liar's paradox,
Cantor's diagonal argument, Russell's paradox, G\"odel's (first) incompleteness
theorem, Tarsky's theorem, undecidability of the Halting problem, Kleene's
fixed point (recursion) theorem and von Neumann's construction of
self-reproducing automata~--- in all these cases the core argument is essentially
the same.

The same trick is used  also in a classical programming challenge: to write a program
that prints its own text. Of course, for every string $s$ it is trivial to write a program $t(s)$
that prints $s$, but how do we get $t(s)=s$? It seems at first that $t(s)$ should
incorporate the string $s$ itself plus some overhead, so how $t(s)$ can be
equal to $s$? However, this first impression is false.  Imagine that our
computational device is a universal Turing machine $U$ where the program is written
in a special read-only layer of the tape. (This means that the tape alphabet is
a Cartesian product of two components, and one of the components is used for
the program and is never changed by $U$.) Then the program can get access to
its own text at any moment, and, in particular, can copy it to the output tape.\footnote{%
   Of course, this looks like cheating: we use some very special
   universal machine as an interpreter of our programs, and this
   makes our task easy. Teachers of programming that are
   seasoned enough may recall the BASIC program
   \begin{center}
   \texttt{10\quad LIST}
   \end{center}
   that indeed prints its own text. However, this trick can be
   generalized enough to show that a self-printing program
   exists in every language.}
Now we explain in more details how to get a self-similar tile set
according to this scheme.

\section{Implementing a given tile set}
\label{implementing}

In this section we show how one can implement a given tile set
$\sigma$, or, better to say, how to construct a tile set $\tau$
that implements some set of macro-tiles that is isomorphic to
$\sigma$.

There are easy ways to do this. Though we cannot let
$\tau=\sigma$ (recall that zoom factor $M$ should be greater
than~$1$), we can do essentially the same for every $M>1$. Let
us extend our ``positive'' example (with one macro-tile and $M^2$
tiles) by superimposing additional colors. Superimposing
two sets of colors means the we consider the Cartesian
product of color sets (so each edge carries a pair of colors).
One set of colors remains the same ($M^2$ colors for $M^2$ pairs
of integers modulo $M$). Let us describe additional (superimposed)
colors.
Internal edges of
each macro-tile should have the same color and this color
should be different for all macro-tiles, so we allocate
$\#\sigma$ colors for that. This gives $\#\sigma$ macro-tiles
that can be put into $1$-$1$-correspondence with $\sigma$-tiles.
It remains to provide correct border colors, and this
is easy to do since
each tile ``knows'' which $\sigma$-tile it simulates (due to the
internal color). In this way we get $M^2\#\sigma$ tiles that
implement the tile set $\sigma$ with zoom factor $M$.

However, this (trivial) simulation is not really useful. Recall
that our goal is to get isomorphic $\sigma$ and $\tau$, and in this
implementation
$\tau$-tiles have more colors that $\sigma$-tiles (and we have more
tiles, too). So we need a more creative encoding of
$\sigma$-colors that makes use of the space available: a side of
a macro-tile has a ``macro-color'' that is a sequence of $M$
tile colors, and we can have a lot of macro-colors in this way.

So let us assume that colors in $\sigma$ are $k$-bit strings for
some $k$. Then the tile set is a subset $S\subset
\mathbb{B}^k\times\mathbb{B}^k
\times\mathbb{B}^k\times\mathbb{B}^k$, i.e., a $4$-ary predicate
on the set $\mathbb{B}^k$ of $k$-bit strings. Assume that $S$ is
presented by a program that computes Boolean value $S(x,y,z,w)$
given four $k$-bit strings $x,y,z,w$. Then we can construct a
tile set $\tau$ as follows.

We start again with a set of $M^2$ tiles from our example and
superimpose additional colors but use them in a more economical
way. Assuming that $k\ll M$, we allocate $k$ places in the
middle of each side of a macro-tile and allow each of them to carry an
additional color bit; then a macro-color represents a $k$-bit string.
Then we need to arrange the internal colors in such a way that
macro-colors ($k$-bit strings) $x$, $y$, $z$ and $w$ can appear
on the four sides of a macro-tile if and only if $S(x,y,z,w)$ is
true.

To achieve this goal, let
us agree that the middle part  (of size, say, $M/2\times M/2$)
in every $M\times M$-macro-tile is
a ``computation zone''. Tiling
rules (for superimposed colors) in this zone guarantee that it
represents a time-space diagram of a computation of some (fixed)
universal Turing machine.
(We assume that time goes up in a vertical direction
and the tape is horizontal.)
It is convenient to assume that
program of this machine is written on a special read-only layer
of the tape (see the discussion in Section~\ref{fixed}).

Outside the computation zone the tiling rules guarantee that
bits are transmitted from the sides to the initial configuration
of a computation.

\begin{figure}[h]
 \begin{center}
\includegraphics[scale=1]{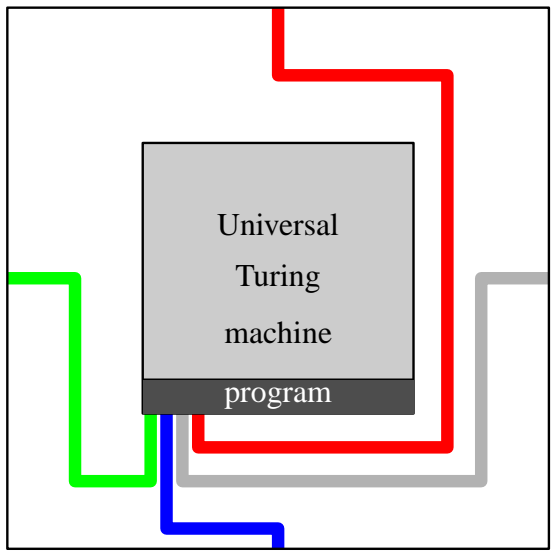}
 \end{center}
\caption{$k$-macro-colors are transmitted to the computation
        zone where they are checked}
\end{figure}

We also require that this machine should accept its
input before running out of time (i.e., less than in $M/2$
steps), otherwise the tiling is impossible.

Note that in this description different parts of a macro-tile
behave differently; this is OK since we start from our example
where each tile ``knows'' its position in a macro-tile (keeps
two integers modulo $M$). So the tiles in the ``wire'' zone know
that they should transmit a bit, the tiles inside the computation
zone know they should obey the local rules for time-space
diagram of the computation, etc.

This construction uses only bounded number of
additional colors since we have fixed the universal Turing
machine (including its alphabet and number of states); we do not need to
increase the number of colors when we increase $M$ and $k$
(though $k$ should be small compared to $M$ to leave enough
space for the wires; we do not give an exact position of the
wires but it is easy to see that if $k/M$ is small enough, there
is enough space for them). So the construction uses $O(M^2)$
colors (and tiles).

\section{A tile set that implements itself}
\label{itself}

Now we come to the crucial point in our argument: can we arrange
things in such a way that the predicate $S$ (i.e., the tile set it
generates) is isomorphic to the set of tiles $\tau$ used to
implement it?

Assume that $k=2\log M +O(1)$; then macro-colors have enough
space to encode the coordinates modulo $M$ plus superimposed
colors (which require $O(1)$ bits for encoding).

Note that many of the rules that define $\tau$ do not depend on
$\sigma$ (i.e., on the predicate $S$). So the program for the
universal Turing machine may start by checking these rules.
It should check that

\begin{itemize}

\item bits that represent coordinates (integers modulo $M$) on
the four sides of a macro-tile are related in the proper way
(left and lower sides have identical coordinates, on the
right/upper side one of the coordinates increases modulo $M$);

\item if the macro-tile is outside computation zone and the
wires, it does not carry additional colors;

\item if the macro-tile is a part of a wire, then it transmits a
bit in a required direction (of course, for this we should fix
the position of the wires by some formulas that are then checked
by a program);

\item if the macro-tile is a part of the computation zone, it
should obey the local rules for the computation zone (bits of the
read-only layer should propagate vertically, bits that encode the
content of the tape and the head of our universal
Turing machine should change
as time increases according to the behavior of this machine, etc.)

\end{itemize}

This guarantees that on the next layer
macro-tiles are grouped into macro-macro-tiles where
bits are transmitted correctly to the computation zone of a macro-macro-tile
and \emph{some} computation of the universal Turing machine
is performed in this zone. But we need more: this computation should be the
same computation that is performed on the macro-tile level
(fixed point!). This is also easy to achieve since in our model
the text of a running program is available to it (recall the we assume
that the program is written in a read-only layer): the program should check
also that \emph{if a macro-tile is in the computation zone, then
the program bit it carries is correct} (program knows the
$x$-coordinate of a macro-tile, so it can go at the
corresponding place of its own tape to find out which program
bit resides in this place).

This sound like some magic, but we hope that our previous
example (a program for the UTM that prints its own text) makes
this trick less magical (indeed, reliable and reusable magic is
called technology).

\section{So what?}

We believe that our proof is rather natural. If von Neumann
lived few years more and were asked about aperiodic tile sets,
he would probably immediately give this argument as a solution.
(He was especially well prepared to it since he used very
similar self-referential tricks to construct a self-reproducing
automata, see \cite{neumann}.) In fact this proof somehow
appeared, though not very explicitly, in P.~G\'acs' papers on
cellular automata~\cite{gacs}; the attempts to understand these
papers were our starting points.

This proof is rather flexible and can be adapted to get many
results usually associated with aperiodic tilings:
undecidability of domino problem (Berger~\cite{berger}),
recursive inseparability of periodic tile sets and inconsistent
tile sets (Gurevich -- Koryakov~\cite{gurevich-koryakov}),
enforcing substitution rules (Mozes~\cite{mozes}) and others
(see~\cite{gurevich,dlt}). But does it give something new?

We believe that indeed there are some applications that hardly
could be achieved by previous arguments. Let us conclude
by mentioning two of
them. First is the construction of \emph{robust} aperiodic tile sets. We can
consider tilings with holes (where no tiles are placed and
therefore no matching rules are checked). A robust aperiodic
tile set should have the following property: if the set of holes
is ``sparse enough'', then tiling still should be far from any
periodic pattern (say, in the sense of Besicovitch distance, i.e.,
the limsup of the fraction of mismatched positions in a centered square
as the size of the square goes to infinity).
The notion of ``sparsity'' should not be too restrictive here;
we guarantee, for example, that a Bernoulli random set with small
enough probability $p$ (each cell belongs to a hole
independently with probability $p$) is sparse.

While the first example (robust aperiodic tile sets) is
rather technical (see~\cite{dlt} for details),
the second is more basic. Let us split all tiles in some tile
set into two classes, say, A- and B-tiles. Then we consider a
fraction of A-tiles in a tiling. If a tile set is not restrictive
(allows many tilings), this fraction could vary from one tiling
to another. For classical aperiodic tilings this fraction is
usually fixed: in a big tiled region the fraction of
A-tiles is close to some limit value, usually an eigenvalue of
an integer matrix (and therefore an algebraic number). The
fixed-point construction allows us to get any computable number.
Here is the formal statement: \emph{for any
computable real $\alpha\in[0,1]$ there exists a tile set $\tau$
divided into $A$- and $B$-tiles such that for any
$\varepsilon>0$ there exists $N$ such that for all $n>N$ the
fraction of A-tiles in any $\tau$-tiling of $n\times n$-square
is between $\alpha-\varepsilon$ and $\alpha+\varepsilon$.}


\begin{thebibliography}{9}

\bibitem{berger}
R.~Berger, The Undecidability of the Domino Problem.
\emph{Mem. Amer. Math. Soc.},
\textbf{66}, 1966.

\bibitem{gurevich}
Egon B\"orger, Erich Gr\"adel, Yuri Gurevich,
\emph{The Classical Decision Problem},  Springer, 1987.
ISBN~3-540-57073-X

\bibitem{intelligencer}
Bruno Durand, Leonid Levin, Alexander Shen, Local Rules and
Global Order, or Aperiodic Tilings, \emph{Mathematical
Intelligencer}, \textbf{27}(1), 64--68 (2005).

\bibitem{dlt}
Bruno Durand, Andrei Romashchenko, Alexander Shen,
Fixed Point and Aperiodic Tilings,
\emph{Developments in Language Theory, 12th International
Conference, DLT 2008, Kyoto, Japan, September 16--19, 2008.
Proceedings}, Springer, Lecture Notes in Computer Science, volume 5257,
2008. ISBN: 978-3-540-85779-2.

\bibitem{gacs}
Peter G\'acs, Reliable Cellular Automata with Self-Organization,
\emph{J. Stat. Phys.}, \textbf{103} (1/2), 45--267, 2001.

\bibitem{grunbaum}
Branko Grunbaum, Geoffrey C. Shephard, \emph{Tilings and Patterns},
W.H.Freeman \& Co, 1986.

\bibitem{gurevich-koryakov}
Yuri Gurevich, Igor Koryakov, A remark ob Berger's paper on the
domino problem, \emph{Siberian Mathematical Journal}, \textbf{13}, 319--321,
1972.

\bibitem{mozes}
Shahar Mozes,
Tilings, Substitution Systems and Dynamical Systems Generated by
Them, \emph{J. Analyse Math.}, \textbf{53}, 139--186, 1989.

\bibitem{neumann}
John von Neumann,
\emph{Theory of Self-reproducing Automata}, edited by A.~Burks,
University of Illinois Press, 1966.

\bibitem{wang}
Hao Wang, Proving theorems by pattern recognition II, \emph{Bell System Technical Journal},
\textbf{40}, 1--42 (1961).
\end{thebibliography}
\end{document}